\documentclass{article}
\usepackage{graphicx} 
\usepackage{amsmath}
\usepackage{amsfonts}
\usepackage{natbib}
\usepackage{url}
\usepackage{pdfpages}

\title{Research Funding as a Decision Problem Under Heavy-Tailed Uncertainty}
\author{Carlos Oscar S. Sorzano, B. Pueche-Granados}
\date{\today}

\begin{document}

\maketitle

\begin{abstract}
Heavy-tailed impact distributions, intrinsic uncertainty, and the high costs of proposal-based peer review increasingly challenge research funding decisions. Using large-scale bibliometric data, we show that past scientific performance provides statistically meaningful, though imperfect, information about future productivity and impact across multiple dimensions. An aggregated, percentile-normalised proxy signal captures this predictive structure robustly across research domains. 

We analyse deterministic and stochastic funding allocation mechanisms under impact-based objectives and find that both converge to highly concentrated allocations that favour a small number of top-performing researchers. To address the limitations of pure exploitation, we introduce a biased lottery framework based on a regularised decision-theoretic objective that explicitly balances exploration and exploitation while accounting for practical funding constraints. Our results suggest that biased lottery mechanisms offer a transparent, efficient, and scalable alternative to conventional peer review in environments characterised by heavy-tailed scientific returns. Additionally, we provide a web application, available at \url{http://scilottery.biocomputingunit.es}, that implements the deterministic allocation method presented in this work.
\end{abstract}

\section{Introduction}

The allocation of public research funds is a central instrument of science policy, intended to maximise the long-term societal return of scientific activity. In most contemporary systems, this allocation relies predominantly on peer review, in which expert panels assess proposals and applicants to identify those most likely to produce impactful research. Despite its widespread adoption and clear merits, peer review operates under severe uncertainty: scientific outcomes are inherently unpredictable, and the eventual impact of funded research often differs substantially from expectations formed at the time of evaluation.

From a policy perspective, the goal of research funding is not merely to sustain activity, but to maximise societal return, broadly understood as the production of knowledge, methods, and discoveries that advance science. Although societal impact is difficult to measure directly and often unfolds over long time scales, bibliometric indicators are commonly used as practical proxies for scientific output and influence within defined fields and time windows \citep{Wilsdon2015, Sinatra2016}. Measures of productivity, citation impact, and the occurrence of highly influential publications capture complementary aspects of scientific performance and remain central---explicitly or implicitly---to many funding decisions.

A natural assumption underlying these practices is that researchers who have performed well in the recent past are, on average, more likely to perform well in the near future, although high-impact contributions can occur at any point in their scientific careers \citep{Sinatra2016}. This intuition motivates both performance-based funding schemes and peer-review judgments informed by publication records. However, it raises two closely related but unresolved questions. First, to what extent are bibliometric indicators computed over a recent period actually predictive of future performance over a comparable horizon? Second, if such predictability exists, how should it be used to allocate limited research funds in a setting where scientific impact is highly skewed, with rare and weakly predictable breakthroughs accounting for a disproportionate share of total influence? The first question has been examined in a growing body of work showing that scientific careers exhibit measurable persistence and cumulative advantage dynamics. Longitudinal analyses of publication and citation records reveal statistically significant temporal correlations between productivity and impact \citep{Petersen2012}, while large-scale studies of career trajectories demonstrate substantial heterogeneity yet identifiable regularities in performance over time \citep{Way2017}. At the same time, research on the predictability of scientific impact indicates that although early signals contain informative structure, individual breakthroughs remain only partially foreseeable due to intrinsic stochasticity and heavy-tailed citation distributions \citep{Wang2013e, Sinatra2016}. These findings suggest that past performance provides a meaningful—but imperfect—basis for forecasting future outcomes, reinforcing the need to distinguish between statistical predictability at the aggregate level and the irreducible uncertainty surrounding exceptional contributions.

In this work, we address these questions by framing research funding as a decision problem under heavy-tailed uncertainty \citep{Wang2013e}. We separate the empirical assessment of predictability from the normative issue of optimal allocation. Using large-scale bibliometric data, we quantify how past performance predicts future outcomes across multiple dimensions and examine the implications of this predictability for funding policies designed to maximize scientific impact. We show that impact-based optimization — whether implemented through deterministic rules or stochastic mechanisms — naturally yields highly concentrated allocations. Building on this result, we introduce a biased lottery framework derived from a regularized decision-theoretic objective that explicitly balances the exploitation of predictive signals with exploration under irreducible uncertainty. 

This approach provides a transparent and scalable alternative to conventional proposal-based peer review \citep{Fang2016, dePeuter2022}, which becomes increasingly unreliable as funding rates decline. At such paylines, panels are effectively asked to make fine-grained distinctions among many highly meritorious proposals, despite strong evidence that reviewers have limited statistical power to accurately stratify applications within the top tier \citep{Fang2016b} or provide consistent scores \citep{Marsh2008, Bornmann2011}. In this regime, small random fluctuations in scoring and the influence of individual reviewers can determine outcomes \citep{Kaplan2008}, magnifying arbitrariness and exacerbating bias related to factors such as seniority, institutional prestige, gender, race, or field proximity \citep{Fang2016, Guthrie2019}.

\section{Methods}

We analyzed large-scale bibliometric data obtained from the OpenAlex database through its public API. The study focuses on individual researchers affiliated with scientifically homogeneous research institutes, selected to minimize heterogeneity in publication and citation practices across disciplines. This approach helps to standardize the analyses independently of the research topic, an alternative to the approach discussed in \citet{Radicchi2008}. All analyses were conducted at the researcher level.

For each researcher, bibliometric indicators were computed over two consecutive, non-overlapping five-year windows: a reference period $[Y-5, Y-1]$ and a future period $[Y, Y+4]$. Within each window, we extracted three complementary indicators of scientific performance: productivity, measured as the number of publications; average citation impact, measured as citations per publication; and maximum citation impact, defined as the highest number of citations received by any single publication. Citations were counted only within the same temporal window as the corresponding publications. Researchers with fewer than two publications in the reference period were excluded to ensure minimally stable performance estimates.

To enable meaningful comparisons across researchers operating in related but non-identical scientific contexts, all indicators were normalised using within-institute percentile ranks. For each institute and time window, raw bibliometric measures were replaced by their empirical percentile ranks and rescaled to the interval $[0,1]$. This normalisation removes scale effects associated with field-specific publication volumes and citation dynamics, ensuring that all indicators are interpreted as relative standing within a homogeneous scientific context.

In addition to individual indicators, we defined an aggregated past-performance signal by averaging percentile-normalised productivity, the average citation impact, and the maximum citation impact over the reference period. This composite proxy provides a compact, field-normalised summary of recent scientific performance and serves as the primary predictive signal in subsequent analyses.

Predictability of future scientific performance was assessed by comparing percentile-normalised indicators measured in the reference period with their counterparts in the coming period. We combined exploratory visualisation, cross-period rank correlations, dimensionality reduction, and rank-based regression models to quantify performance stability over time. Rank-based methods were used throughout to focus on relative ordering rather than absolute magnitudes, ensuring robustness to heavy-tailed distributions and extreme observations and reflecting the comparative nature of evaluation and funding decisions. Full methodological details and additional analyses are provided in the Supplementary Material.

\subsection{Deterministic allocation model}
\label{sec:deterministic_allocation}

We introduce a deterministic funding allocation rule that combines exploitation of predictive performance signals with an explicit exploration component that accounts for irreducible uncertainty and heavy-tailed scientific returns. 

Let $B$ denote the total available budget to be distributed among $N$ eligible researchers. Each researcher $i$ is assigned a non-negative performance score $s_i$, derived from reference-period indicators and predictive modelling, with larger values indicating higher expected future performance. An allocation $\mathbf{b}=(b_1,\dots,b_N)$ satisfies $\sum_i b_i = B$ and $b_i \ge 0$.

The allocation rule can be interpreted as an approximation to a utility-maximisation problem that balances predictable returns with the option value of rare, high-impact outcomes under heavy-tailed uncertainty,
\[
U(\mathbf{b}) = (1-\alpha)\sum_{i=1}^N b_i s_i + \alpha\,\mathrm{Tail}(\mathbf{b}),
\]
where $\alpha\in[0,1]$ controls the trade-off between exploitation of predictive signals and preservation of exploration. The first term corresponds to expected return under the predictive model, while the second captures the value of maintaining broad support to increase exposure to unpredictable but potentially transformative contributions. In the deterministic setting, this tail component is approximated through diversification rather than explicit stochastic modelling.

Operationally, the total budget $B$ is split into an exploration component
$B_{\mathrm{explore}}=\alpha B$ and an exploitation component
$B_{\mathrm{exploit}}=(1-\alpha)B$.
The exploration budget is distributed broadly across researchers while allowing
a controlled dependence on predicted performance,
\[
b^{\mathrm{explore}}_i
=
B_{\mathrm{explore}}
\left(
\lambda \frac{1}{N}
+
(1-\lambda)\frac{s_i}{\sum_{j=1}^N s_j}
\right),
\]
where $\lambda\in[0,1]$ interpolates between a uniform allocation
($\lambda=1$) and one proportional to the performance scores $s_i$ ($\lambda=0$).
Larger values of $\lambda$ preserve broad participation while retaining weak
performance sensitivity.

In contrast, the exploitation budget is allocated by concentrating resources on higher-scoring researchers using a parametric concentration rule. In particular, the exploitation component takes the form
\[
b_i^{\mathrm{exploit}} = B_{\mathrm{exploit}}
\frac{s_i^{\gamma}}{\sum_{j=1}^N s_j^{\gamma}},
\]
where $\gamma>0$ controls the degree of concentration, ranging from egalitarian ($\gamma<1$) to strongly concentrated allocations ($\gamma\gg 1$). The total allocation to each researcher is given by $b_i = b_i^{\mathrm{explore}} + b_i^{\mathrm{exploit}}$.

The resulting policy can be interpreted as a portfolio rule: a fraction $\alpha$ of the budget maintains exposure to unpredictable, high-impact outcomes, while the remaining fraction is invested based on predicted performance, with a tunable concentration. Optional lower and upper bounds on individual allocations can be imposed to reflect policy or operational constraints.

The allocation rule is governed by a small set of interpretable hyperparameters controlling exploration, uniformity, and concentration. These parameters can be fixed by policy design or optimised empirically via backtesting on historical data. Full mathematical details, alternative formulations, and implementation specifics are provided in the Supplementary Material.

\subsection{Stochastic allocation model}
\label{sec:stochastic_allocation}

The deterministic allocation rule provides a transparent baseline but treats uncertainty only implicitly. Because scientific returns are heavy-tailed and only partially predictable, rare and disproportionate contributions play a dominant role in aggregate impact. We therefore introduce a stochastic allocation mechanism that explicitly incorporates irreducible uncertainty through randomised selection, while remaining grounded in standard decision-theoretic principles.

Let $s_i \ge 0$ denote the predicted performance score of researcher $i$. In the stochastic framework, the decision variable is a probability distribution $p=(p_1,\dots,p_N)$ over researchers, where $p_i$ represents the probability that researcher $i$ is selected for funding. We consider a regularised expected-utility objective of the form
\[
\max_{p}
\;
(1-\alpha)\sum_{i=1}^N p_i s_i
-
\alpha\,\tau\,\mathrm{KL}\!\left(p\,\|\,u\right),
\]
subject to $\sum_i p_i = 1$ and $p_i \ge 0$, where $u$ denotes the uniform distribution, $\mathrm{KL}(\cdot\|\cdot)$ is the Kullback--Leibler divergence, $\alpha\in[0,1]$ controls the exploitation--exploration trade-off, and $\tau>0$ is a temperature parameter. The first term favors concentration on high-scoring researchers, while the regularization term penalizes excessive concentration and preserves exposure to unpredictable outcomes.

This optimization admits a closed-form solution given by a Gibbs (softmax) distribution,
\[
p_i
=
\frac{\exp\!\left(\dfrac{(1-\alpha)s_i}{\alpha\tau}\right)}
{\sum_{j=1}^N \exp\!\left(\dfrac{(1-\alpha)s_j}{\alpha\tau}\right)}.
\]
As $\tau \to \infty$, the policy converges to uniform random selection, whereas as $\tau \to 0^+$ it approaches a purely exploitative rule concentrated on the highest-scoring researchers. Intermediate values yield a principled mixture of exploitation and exploration. Importantly, this lottery is not \emph{ad hoc}: it arises as the unique optimizer of a regularized expected-utility problem closely related to maximum-entropy decision making \citep{Ortega2013}.

To reflect the discrete nature of real funding calls, exactly $K$ researchers are selected in each round by sampling without replacement according to the probabilities $\{p_i\}$. Following selection, funding is allocated in two stages: a fixed seed grant ensures a minimum level of support for each selected researcher. At the same time, the remaining budget is distributed conditionally on selection using a concentration rule analogous to the deterministic model. This separation preserves stochastic exploration at the selection stage while exploiting predictive information during allocation. Full mathematical details, implementation choices, and parameter selection procedures are provided in the Supplementary Material.

\section{Results}

Our empirical analysis is based on a large cohort of researchers affiliated with scientifically homogeneous research institutes, constructed to enable meaningful within-field comparisons. After applying inclusion criteria and ensuring minimal research activity, the final dataset comprises approximately 4,400 researchers across 20 institutions. Full details on data collection, cohort construction, and normalisation procedures are provided in the Supplementary Material.

Across all institutions and performance dimensions, scientific output and impact exhibit substantial heterogeneity and heavy-tailed behaviour. While a small fraction of researchers accounts for a disproportionate share of publications and citations, relative scientific performance displays partial persistence across consecutive five-year periods. In particular, productivity shows strong temporal stability, whereas citation-based indicators—especially average citation impact—are markedly more variable, reflecting the stochastic nature of scientific recognition.

Despite this variability, past performance contains statistically meaningful information about future outcomes. Rank-based analyses reveal consistently positive cross-period associations for productivity, citation impact, and exceptionally influential contributions, although the strength of the predictive relationships varies across indicators. Dimensionality reduction and collinearity analyses further show that most of the informative content of past performance can be captured by a low-dimensional representation, with sustained productivity emerging as the most stable and informative individual signal. Detailed exploratory analyses, correlation matrices, principal component decompositions, and regression results are reported in the Supplementary Material.

To obtain a compact and robust summary of past scientific performance, we define an aggregated proxy signal, $\texttt{avgPerc1}$, as the average of percentile-normalised productivity, average citation impact, and maximum citation impact measured in the reference period. This composite indicator integrates sustained output, typical influence, and exceptional contributions into a single, field-normalised score.

\begin{figure}[htbp]
\centering
\includegraphics[width=\textwidth]{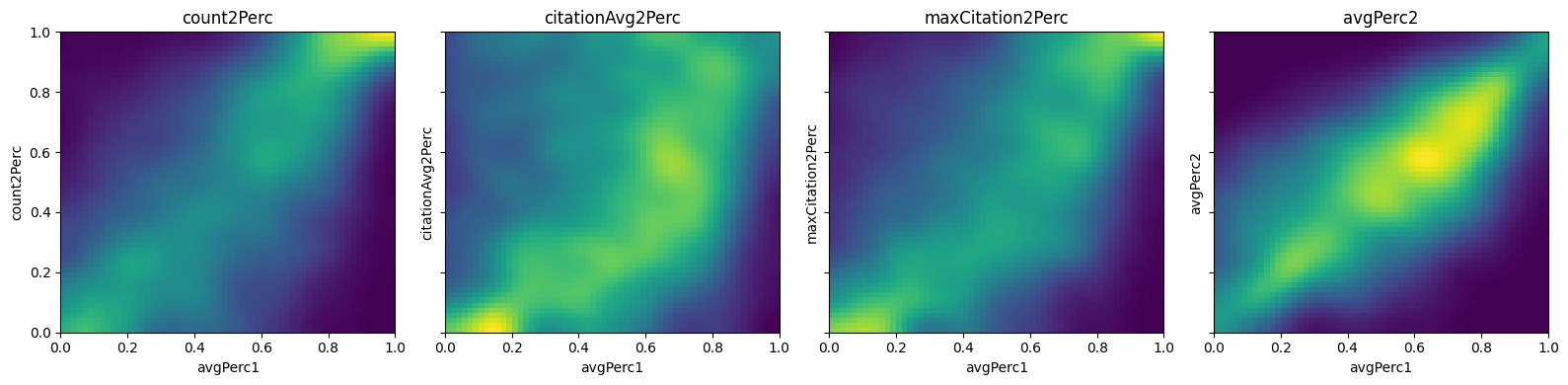}
\caption{Smoothed joint distributions of percentile-normalised future outcomes as a function of aggregated past performance $\texttt{avgPerc1}$. Each panel shows a kernel-smoothed two-dimensional histogram for a different outcome in the future period: productivity, average citation impact, maximum citation impact, and the aggregated outcome $\texttt{avgPerc2}$. The concentration of probability mass along a monotonic diagonal in all panels indicates that $\texttt{avgPerc1}$ is a strong predictor across all dimensions of future performance, despite substantial dispersion reflecting irreducible uncertainty.}
\label{fig:avgperc1_predictor}
\end{figure}

Figure~\ref{fig:avgperc1_predictor} shows that this aggregated proxy signal is a strong predictor of future scientific performance across all outcome dimensions considered. Higher values of $\texttt{avgPerc1}$ are associated with systematically higher median and upper-quantile outcomes in future productivity, citation impact, and aggregated performance. At the same time, substantial dispersion persists at all levels of past performance, particularly for citation-based measures, indicating that even strong aggregation cannot eliminate irreducible uncertainty or heavy-tailed effects at the individual level.

Taken together, these results establish two key empirical facts that motivate the allocation models introduced below. First, recent scientific performance provides a statistically informative but imperfect signal of future outcomes. Second, exceptional contributions remain only weakly predictable, even when conditioning on strong past performance. Any funding mechanism aimed at maximising scientific impact must therefore balance exploitation of predictable signals with explicit accommodation of uncertainty and tail risk.

\subsection{Optimisation of deterministic and stochastic allocation policies}

We evaluated the performance of both deterministic and stochastic allocation rules by optimising their parameters against realised scientific outcomes in the future period. In both cases, the objective function was defined as
\[
U = \sum_{i=1}^{N} b_i\, s_i,
\]
where $b_i$ denotes the fraction of the total budget allocated to researcher $i$ and $s_i$ is the percentile-normalised composite performance indicator in the future period ($\texttt{avgPerc2}$). The total budget was fixed to $B=1$, so that allocations can be interpreted as budget shares.

For the deterministic policy, we performed a grid search over the hyperparameters $(\alpha,\lambda,\gamma)$, which control, respectively, the fraction of the exploration budget, the uniformity of the exploration component, and the concentration of the exploitation component. Across the explored parameter ranges, optimisation consistently selected $\alpha=0$ and $\lambda=0$, corresponding to a purely exploitative policy without an explicit exploration component. Under this configuration, utility increased monotonically with the concentration parameter $\gamma$, reaching its maximum at the largest value considered. This indicates that greater resource allocation to high-scoring researchers yields higher realised utility when performance is evaluated solely on future impact.

\begin{figure}[htbp] \centering \includegraphics[width=0.9\textwidth]{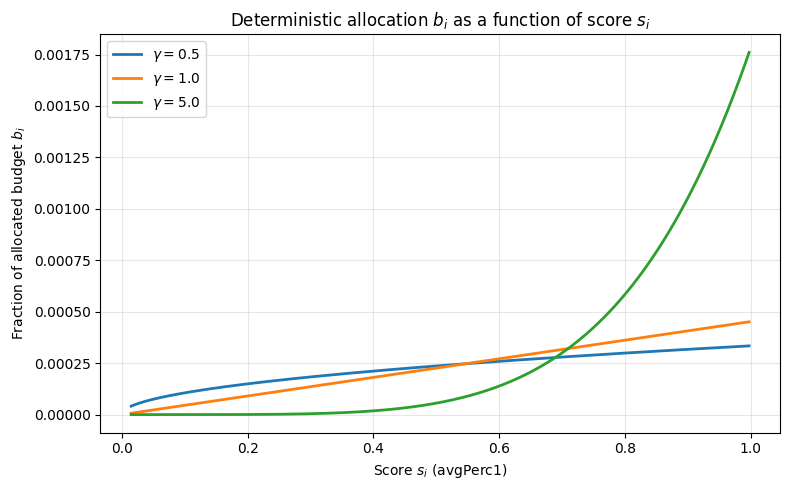} \caption{Deterministic allocation curves showing the fraction of the total budget $b_i$ assigned to each researcher as a function of the reference-period score $s_i$ for $\alpha=\lambda=0$ and different values of the concentration parameter $\gamma$. Increasing $\gamma$ leads to progressively more concentrated allocations, strongly favouring researchers with the highest predicted performance.} \label{fig:deterministic_allocation_curves} \end{figure}

Figure~\ref{fig:deterministic_allocation_curves} illustrates the resulting deterministic allocation rule for $\alpha=\lambda=0$ and representative values of $\gamma$. As $\gamma$ increases, allocations become increasingly convex and concentrated, strongly favouring researchers with the highest predicted performance.

We conducted an analogous optimisation for the stochastic allocation mechanism by maximising expected utility, estimated by averaging outcomes over multiple independent realisations of the lottery. Despite its explicit treatment of uncertainty, the optimal stochastic policy also collapses to an extreme-concentration regime. The optimal configuration selects a tiny subset of researchers ($K=5$ out of approximately 4,400), assigns a minimal seed grant, and converges to $\alpha=0$, corresponding to a purely exploitative selection policy. Conditional on selection, the remaining budget is distributed relatively evenly among the selected researchers.

These results show that both deterministic and stochastic allocation mechanisms converge toward highly concentrated, exploitative solutions when evaluated solely on average future performance. This convergence is not an artefact of the allocation rule, but a direct consequence of the objective function: under heavy-tailed outcome distributions and partial predictability, maximising expected impact systematically favours concentrating resources on a small number of top-performing researchers. Exploration and diversification do not arise endogenously from impact-only optimisation, motivating their explicit inclusion as normative policy objectives rather than emergent outcomes.

\section{Discussion}

This study reveals a fundamental tension in research funding: future scientific performance exhibits substantial variability and heavy-tailed behaviour, yet past performance provides a statistically meaningful—though imperfect—predictive signal. Citation-based measures, in particular, display pronounced dispersion, reflecting the inherently stochastic nature of scientific influence. At the same time, researchers with stronger recent performance tend to achieve higher productivity and impact in subsequent periods. An aggregated proxy signal combining productivity, average citation impact, and exceptional contributions captures this predictive structure more robustly than any single indicator, providing a compact summary of expected future performance under heavy-tailed uncertainty.

A second central finding is that both deterministic and stochastic allocation mechanisms converge toward extreme concentration when evaluated solely on expected future impact. Regardless of whether exploration is implemented through deterministic spreading or randomised selection, impact-based optimisation systematically favours allocating resources to a small number of top-performing researchers. This outcome is not an artefact of the allocation rules, but a direct consequence of the objective function: under partial predictability and heavy-tailed returns, concentrating resources on established excellence maximises expected output. In fact, this tendency is consistent with our empirical observation that past performance is informative about future performance: researchers who rank highly in productivity and impact in one period tend, on average, to remain among the most productive in the next. At the same time, this impact-driven concentration should be interpreted in light of the broader science-policy debate reviewed by \citet{Aagaard2020}, who synthesises both the proposed benefits of concentrating funding and the arguments for dispersal across a wider population of researchers. Importantly, these perspectives are not necessarily in opposition. The concentration that emerges from our framework is not concentration ``per se'', but concentration in those researchers who have already demonstrated exceptional performance. Moreover, many of the arguments favouring dispersal in the literature are motivated by objectives that are not reducible to short-term bibliometric performance—such as diversity, renewal, resilience, fairness, and the mitigation of cumulative advantage—rather than by evidence that dispersed allocations strictly maximise expected impact. Where performance-based evidence is available, \citet{Aagaard2020} highlights diminishing marginal returns to additional funding beyond a certain point; however, the same literature also suggests that, before reaching that saturation point, additional resources can generate steep gains in output and influence. Taken together, these findings suggest that impact-only optimisation naturally drives concentrated allocations, while the case for dispersion typically rests on additional policy goals that must be introduced explicitly into the objective function rather than expected to arise endogenously from performance maximisation alone.

These results have direct implications for the design of research funding systems. Current peer-review–based selection mechanisms combine subjective assessments of past performance with speculative evaluations of proposed future work, under conditions of limited expertise, time pressure, and intrinsic uncertainty. Several studies have highlighted problems with the system \citep{Marsh2008, Bornmann2011}. Even in contexts where research has already been completed and rigorously reviewed, predicting impact is difficult; forecasting the success of a proposed study is even more uncertain. This structural limitation contributes to high evaluation costs \citep{Gross2019}, limited transparency, and widespread dissatisfaction among applicants \citep{Vaesen2017, Schweiger2024}.

Against this background, our findings support alternative funding models that treat uncertainty not as a flaw to be eliminated, but as a structural feature of curiosity-driven research that must be managed through institutional design. Biased lottery schemes based on transparent, interpretable indicators can provide a scalable way to balance the benefits of concentrating resources on strong candidates with the need to preserve diversity, renewal, and openness to serendipitous discoveries. By combining evidence-informed prioritisation with a controlled degree of randomness, such mechanisms align funding decisions with the explore–exploit trade-off: they exploit signals that correlate with future performance while maintaining support pathways for unconventional lines of work that cannot be reliably forecasted \textit{ex ante}. At the same time, they respond to well-known systemic tensions in contemporary funding regimes—such as path dependence (the Matthew effect), disparities across regions and demographic groups, and the opportunity costs associated with proposal writing and grant administration—by making policy goals explicit and by enabling empirical comparison of alternative allocation models rather than assuming a single optimal scheme \citep{Gigerenzer2025}. Our framework occupies an intermediate position between fully proposal-based peer review and the fully egalitarian distribution model studied in \citet{Vaesen2017}.

\section{Conclusions}

We examined research funding allocation from a data-driven, decision-theoretic perspective. Using large-scale bibliometric data, we showed that scientific performance exhibits substantial variability and heavy-tailed behaviour, yet retains statistically meaningful predictability across consecutive periods. An aggregated, percentile-normalised proxy signal captures this structure robustly and provides a compact summary of expected future performance across multiple dimensions.

When funding allocation is optimised solely for future impact, both deterministic and stochastic mechanisms converge toward highly concentrated solutions that favour a small number of top-performing researchers. This convergence reflects the combination of partial predictability and heavy-tailed returns. It highlights a fundamental limitation of impact-only optimisation: exploration, diversity, and generational renewal do not emerge endogenously and must be incorporated explicitly as policy objectives.

To address this limitation, we introduced a stochastic allocation framework based on biased lotteries derived from a regularised expected-utility principle. This approach provides a principled way to balance exploitation of predictive signals with exposure to irreducible uncertainty, while offering transparent control over concentration and exploration.

Overall, our results indicate that biased lottery mechanisms are not a relaxation of rigour, but a pragmatic response to uncertainty, heavy-tailed outcomes, and the costs of proposal-based peer review. By complementing or partially replacing subjective evaluation with transparent, data-informed selection rules, funding systems can improve scalability, reduce administrative burden, and allocate a greater share of resources directly to research activity.

\bibliography{bibliography}
\bibliographystyle{apalike}



\section*{Supplementary material}

# Supplementary Material

April 13, 2026

## 1 Methods

### 1.1 Data collection

All bibliometric data used in this study were obtained from the OpenAlex database through its public application programming interface (API). OpenAlex provides structured, openly accessible metadata on scientific publications, authors, institutions, and citation relationships, making it suitable for large-scale and reproducible analyses of scientific activity.

We focused on researchers affiliated with selected research institutions, identified by their OpenAlex institution identifiers. Institutions were chosen to be scientifically homogeneous, meaning they share a well-defined thematic or disciplinary focus. This criterion excludes large comprehensive universities, which aggregate researchers from widely disparate fields with fundamentally different publication, collaboration, and citation practices. Instead, we considered focused research institutes or centers, where researchers operate within a common scientific domain, even if multiple closely related subfields are represented. This choice reduces field-driven heterogeneity and allows for more meaningful comparisons across individuals.

For each institution, we identified researchers whose current or most recent affiliation in OpenAlex corresponded to that institution and whose affiliation history indicated active association during the period of interest. All analyses were conducted at the individual researcher level.

For each researcher $i$, we collected all scientific works indexed in OpenAlex that were published within two non-overlapping five-year windows. The first window, hereafter referred to as the reference period, spans the years $[Y-5, Y-1]$. The second window, the future period, spans the years $[Y, Y+4]$. The year $Y$ thus defines the temporal boundary between retrospective measurement and prospective evaluation. These windows were chosen as a compromise between temporal resolution and statistical robustness, and reflect common evaluation horizons used in research assessment exercises.

Within each period, we extracted the complete list of works authored by each researcher and computed three bibliometric indicators. First, the number of works published during the period serves as a measure of research output. Second, the number of citations received by those works within the same period

captures the contemporaneous scientific influence of the production. Third, the maximum number of citations received by any single work within the period is intended to capture the occurrence of highly influential contributions. Citations were counted only if the citing works were published within the same temporal window as the cited works, ensuring that productivity and impact are measured consistently within each period.

Researchers with zero or one publication during the reference period were excluded from the analysis. This filtering step serves two purposes. First, it removes cases where bibliometric indicators are dominated by statistical noise rather than sustained research activity. Second, it restricts the analysis to researchers with a minimum level of observable output, for whom comparisons across periods are meaningful. The exclusion criterion was applied uniformly across all institutions.

## 1.2 Data preprocessing and normalization

To enable meaningful comparisons across researchers operating in different scientific contexts, we applied a normalization procedure designed to mitigate residual field-dependent effects. Even within thematically focused research institutes, systematic differences remain in publication rates, collaboration patterns, and citation dynamics across subfields. Direct comparisons based on raw bibliometric indicators may therefore be misleading.

To address this issue, all bibliometric measures were transformed into within-group percentiles. In the present study, the grouping unit is the research institute, chosen for its thematic coherence and relative disciplinary homogeneity. Specifically, for each institute and each time window, the raw indicators — namely, the number of publications, the average number of citations per publication, and the maximum number of citations received by any publication — were replaced by their empirical percentile ranks computed over the population of researchers affiliated with that institute.

Percentiles were linearly rescaled to lie in the interval $[0, 1]$, such that a value of 1 corresponds to the highest-performing researcher within a given group and metric, and a value of 0 corresponds to the lowest-performing researcher. Intermediate values reflect relative standing within the same scientific context. This transformation ensures that, by construction, the best and worst performers are consistently identified, independently of absolute differences in publication volume or citation intensity.

Although the normalization is implemented at the research institute level here, the approach is more general. Any grouping that defines a scientifically homogeneous domain—such as a narrowly defined discipline, subfield, or evaluation panel—can serve as the reference set for percentile computation. The essential requirement is that researchers within the group share comparable publication and citation practices. Under these conditions, percentile-based normalization provides a flexible and extensible framework for harmonizing bibliometric indicators across heterogeneous scientific domains.

By construction, this normalization removes scale effects associated with

field-specific practices and allows the resulting indicators to be interpreted uniformly across groups. A normalized value of 1 always denotes top performance and a value of 0 denotes bottom performance, irrespective of the underlying scientific domain. This percentile-based representation forms the basis for all subsequent analyses, including predictability assessment and the formulation of funding allocation strategies.

## 1.3 Predictability analysis

The objective of the predictability analysis is to assess the extent to which bibliometric indicators computed over the reference period provide informative signals about future scientific performance. Given the heavy-tailed nature of scientific output and impact, this analysis combines exploratory, correlational, and regression-based approaches, with particular emphasis on rank-based methods.

**Exploratory data analysis.** We begin with an exploratory data analysis to visualize the relationship between performance indicators measured in the reference period and their counterparts in the future period. Scatter plots are used to examine pairwise associations between normalized indicators across periods, allowing the identification of broad trends, nonlinearity, and heteroskedasticity. In addition, researchers are grouped into deciles according to their percentile rank in the reference period, and boxplots of future-period performance are constructed for each decile. This representation highlights systematic shifts in central tendency and dispersion across performance strata, and provides an intuitive visualization of predictability at different levels of past performance.

**Cross-period correlation analysis.** To quantify the stability of bibliometric indicators over time, we compute cross-period correlations between the normalized ranks obtained in the reference period and those obtained in the future period. Correlation analyses are performed separately for productivity (number of publications), average citation impact, and maximum citation impact. Rank-based correlations are emphasized to focus on relative ordering rather than absolute magnitudes and to reduce sensitivity to extreme values. These correlations provide a first quantitative estimate of the extent to which past performance predicts future standing within a homogeneous scientific context.

**Dimensionality reduction and collinearity assessment.** Bibliometric indicators are not independent by construction: productivity, average impact, and maximum impact may capture overlapping aspects of scientific activity. To assess redundancy and identify a parsimonious set of predictive variables, we perform a collinearity analysis using the Variance Inflation Factor (VIF). Variables with high VIF values indicate strong multicollinearity and limited marginal informational content.

In parallel, we apply Principal Component Analysis (PCA) to the set of normalized reference-period indicators. PCA provides an orthogonal decompo-

sition of the variance structure and allows us to determine how many latent components are required to capture most of the variability in past performance. Together, VIF and PCA analyses inform the minimal dimensional representation needed to summarize historical scientific performance without substantial loss of information.

**Rank-based regression analysis.** To further characterize predictability, we model future performance ranks as a function of reference-period ranks using rank-based regression models of the form

$$\text{rank}(y_i) = \beta_0 + \beta_1 \text{rank}(x_{i1}) + \beta_2 \text{rank}(x_{i2}) + \varepsilon_i,$$

where $y_i$ denotes a future-period indicator (e.g., productivity or impact), and $x_{i1}, x_{i2}$ denote selected reference-period indicators. All variables entering the regression are expressed as normalized percentile ranks.

Rank-based regression offers several advantages over regression on raw bibliometric measures. First, it is inherently robust to outliers, which are ubiquitous in citation-based indicators due to their heavy-tailed distributions. Second, it avoids assumptions about linearity or homoscedasticity in the original measurement scale, focusing instead on monotonic relationships between variables. Third, rank regression yields coefficients directly interpretable as relative standing within a cohort, which aligns naturally with evaluation and funding decisions that are typically comparative rather than absolute. Finally, by operating on normalized ranks, the regression framework remains invariant to field-specific scaling effects and facilitates comparisons across homogeneous scientific domains.

## 1.4 Hybrid deterministic budget allocation: exploitation with explicit exploration mass

We first introduce a deterministic allocation rule that combines (i) exploitation of predictive signals about future performance and (ii) an explicit exploration component reflecting the possibility that highly influential contributions may arise from any researcher, including those with modest predicted performance. This rule serves as a baseline decision policy before introducing explicitly stochastic or lottery-based mechanisms.

**Scores and objective.** Let $B$ denote the total available budget to be distributed among $N$ eligible researchers. Each researcher $i$ is assigned a non-negative performance score $s_i$, derived from reference-period indicators and predictive modeling. Higher values of $s_i$ indicate higher expected future performance. Crucially, $s_i$ need not be a percentile rank: differences between adjacent researchers may be non-uniform and may encode heterogeneous expected marginal returns.

An entirely exploitative policy would allocate funds to maximize expected return given the scores. However, because scientific returns are heavy-tailed,

4

such a policy can be suboptimal if it concentrates resources too aggressively, thereby reducing the probability of funding unexpected breakthroughs. We therefore introduce a hybrid policy that deterministically allocates a fraction of the budget to broad coverage (exploration) while allocating the remainder to score-driven concentration (exploitation).

**Exploitation and exploration.** The trade-off between exploitation and exploration is a central concept in sequential decision-making and reinforcement learning. In that setting, a greedy policy exploits actions with currently high estimated reward. In contrast, exploratory policies deliberately allocate probability mass to actions that are less certain but may yield rare high rewards or improve future knowledge. In our context, exploitation corresponds to concentrating funds on researchers with higher predicted performance $s_i$. In contrast, exploration corresponds to allocating non-negligible resources to a broad set of researchers to preserve the option value of unforeseen, high-impact outcomes.

**Utility function and tail risk.** The deterministic allocation rule introduced above can be understood as an approximate solution to a utility-maximization problem that explicitly balances predictable returns with the option value of rare, high-impact outcomes. We model the planner's objective as a convex combination of two components:

$$\max_{\mathbf{b}} U(\mathbf{b}) = (1 - \alpha) \sum_{i=1}^{N} b_i \, s_i \; + \; \alpha \, \text{Tail}(\mathbf{b}),$$

subject to $\sum_i b_i = B$ and $b_i \geq 0$. The first term represents the expected return under the assumption that the predictive scores $s_i$ correctly summarize future performance; it favors concentrating resources on high-scoring researchers and corresponds to pure exploitation. The second term captures the value of rare but potentially transformative scientific breakthroughs, which are not reliably predictable from past indicators and are therefore modeled as a function of the allocation vector $\mathbf{b}$ rather than solely the scores.

Importantly, $\text{Tail}(\mathbf{b})$ is not assumed to be a deterministic or directly observable quantity. Instead, it represents the option value created by maintaining sufficiently broad support across researchers, thereby increasing the probability that at least one funded project yields an exceptionally high return. Under heavy-tailed outcome distributions, such tail events may dominate aggregate impact despite their low probability. In the deterministic setting considered here, this tail component is approximated by enforcing diversification in the allocation, rather than by explicitly modeling stochastic outcomes. The parameter $\alpha \in [0, 1]$ thus controls the trade-off between exploitation of predictable performance and preservation of option value under irreducible uncertainty.

This formulation mirrors analogous objectives in reinforcement learning and portfolio optimization, where agents trade off expected reward against risk or uncertainty-driven bonuses. The deterministic hybrid allocation rule introduced

below provides a tractable parametric approximation to this objective, whereas the subsequent lottery-based allocation model explicitly treats the tail term.

**Budget decomposition.** We decompose the budget into an exploration component and an exploitation component:

$$B_{\text{explore}} = \alpha B, \qquad B_{\text{exploit}} = (1-\alpha)B,$$

where $\alpha \in [0,1]$ controls the exploration–exploitation balance. Larger $\alpha$ yields broader coverage; smaller $\alpha$ yields stronger concentration on predicted performance.

**Exploration allocation (baseline plus weak score tilt).** The exploration budget is distributed to preserve broad support while allowing a controlled, weak dependence on predicted performance:

$$b_i^{\text{explore}} = B_{\text{explore}} \left( \lambda \frac{1}{N} + (1-\lambda) \frac{s_i}{\sum_{j=1}^{N} s_j} \right),$$

where $\lambda \in [0,1]$ controls the degree of uniformity. When $\lambda = 1$, the exploration mass is purely uniform and all researchers receive the same exploration allocation. When $\lambda = 0$, the exploration mass becomes fully proportional to $s_i$. In practice, values $\lambda$ closer to one yield a policy that maintains many "lottery tickets" alive while still reflecting predictive information.

**Exploitation allocation (score concentration).** The exploitation budget is distributed by concentrating on higher scores. To control how sharply the allocation concentrates among top scores, we use a power transform:

$$b_i^{\text{exploit}} = B_{\text{exploit}} \frac{s_i^\gamma}{\sum_{j=1}^{N} s_j^\gamma}, \tag{1}$$

where $\gamma > 0$ is a concentration parameter. For $\gamma = 1$, allocation is proportional to $s_i$. For $\gamma > 1$, funding is increasingly concentrated among high-scoring researchers; for $\gamma < 1$, it is more egalitarian.

**Total allocation rule.** The resulting deterministic allocation to researcher $i$ is the sum of the two components:

$$b_i = b_i^{\text{explore}} + b_i^{\text{exploit}}.$$

This policy is interpretable as a portfolio rule: a fraction $\alpha$ of the budget is assigned to maintaining broad exposure to unpredictable high-impact events, while the remaining fraction $(1-\alpha)$ is invested according to predicted performance, with adjustable concentration via $\gamma$.

**Optional minimum and maximum allocations.** Many funding programs impose practical or policy-driven bounds on individual allocations. We incorporate optional constraints

$$b_{\min} \leq b_i \leq b_{\max},$$

where $b_{\min} \geq 0$ ensures a minimum viable level of funding and $b_{\max}$ prevents extreme concentration. When imposed, these bounds are enforced by clipping allocations that fall outside the interval and redistributing any budget surplus or deficit among the remaining eligible researchers while preserving $\sum_i b_i = B$. When $b_{\min} = 0$ and $b_{\max} = +\infty$, the closed-form allocation above applies directly.

**Selection of hyperparameters.** The policy is controlled by $(\alpha, \lambda, \gamma)$, each with a straightforward interpretation: $\alpha$ sets the explore/exploit split, $\lambda$ sets the uniformity of the exploration mass, and $\gamma$ sets the concentration of exploitation. These hyperparameters can be chosen (i) by policy design constraints (e.g., mandated broad coverage) or (ii) empirically via historical backtesting. In the latter case, we simulate allocations using past reference periods and evaluate realized outcomes in the subsequent periods using one or more impact objectives (e.g., total impact, mean impact, or tail-focused metrics such as the maximum impact among funded researchers). The hyperparameters are then selected to optimize the chosen objective under a fixed budget constraint.

This hybrid deterministic policy provides a transparent baseline that explicitly encodes both predictable returns and option value under heavy-tailed uncertainty. In the following subsection, we introduce a lottery-based mechanism that implements exploration through randomized selection rather than deterministic spreading.

## 1.5 Stochastic budget allocation under heavy-tailed uncertainty

The deterministic allocation rule introduced in the previous subsection provides a transparent baseline that exploits predictive information about future scientific performance. However, it treats uncertainty only implicitly and assumes that predicted scores fully summarize future outcomes. The empirical evidence presented above indicates that this assumption is violated: scientific returns are heavy-tailed and only partially predictable, with rare, disproportionate contributions playing a dominant role. In this subsection, we introduce a stochastic allocation mechanism that explicitly incorporates irreducible uncertainty through randomized selection, while remaining firmly grounded in established decision-theoretic principles.

**Motivation and decision-theoretic context.** Randomized decision rules are a standard solution concept in environments characterized by uncertainty,

7

incomplete information, or non-linear payoffs. In game theory, mixed strategies are fundamental and unavoidable whenever pure strategies are suboptimal. In economics and mechanism design, stochastic allocation mechanisms are routinely employed to balance efficiency, robustness, and incentive compatibility. In reinforcement learning and optimal control, stochastic policies arise naturally as optimal solutions under partial observability or heavy-tailed reward distributions.

Research funding shares these structural features. While past performance provides informative signals, the occurrence of exceptionally influential scientific contributions is inherently uncertain and weakly predictable. Under such conditions, purely deterministic allocation rules may overly concentrate resources on predictable returns and reduce exposure to low-probability, high-impact outcomes. A stochastic allocation rule provides a principled way to balance exploitation of predictive signals with exploration under epistemic uncertainty.

**Regularized objective function.** Let $s_i \geq 0$ denote the predicted performance score of researcher $i$, with higher values indicating higher expected future impact. In the stochastic framework, the decision variable is a probability distribution $p = (p_1, \ldots, p_N)$ over researchers, where $p_i$ represents the probability that researcher $i$ is selected for funding.

We consider the following regularized optimization problem:

$$\max_p \ (1-\alpha) \sum_{i=1}^{N} p_i \, s_i \ - \ \alpha \, \tau \, \mathrm{KL}(p \, \| \, u),$$

subject to

$$\sum_{i=1}^{N} p_i = 1, \qquad p_i \geq 0 \ \ \forall i,$$

where $u$ denotes the uniform distribution over researchers, $\mathrm{KL}(p\|u)$ is the Kullback–Leibler divergence, $\alpha \in [0,1]$ controls the trade-off between exploitation and exploration, and $\tau > 0$ is a temperature parameter governing the strength of regularization.

The first term corresponds to expected return under the predictive model and favors concentration on high-scoring researchers. The second term penalizes deviations from uniform selection, thereby encoding epistemic humility and preserving exposure to unpredictable outcomes. This formulation is standard in maximum-entropy decision making and risk-sensitive control.

**Optimal stochastic policy.** The above optimization problem admits a closed-form solution. The optimal selection probabilities are given by the Gibbs (softmax) distribution

$$p_i = \frac{\exp\left(\dfrac{(1-\alpha)s_i}{\alpha \tau}\right)}{\sum_{j=1}^{N} \exp\left(\dfrac{(1-\alpha)s_j}{\alpha \tau}\right)}.$$

This stochastic policy has a straightforward interpretation. As $\tau \to \infty$, the distribution converges to the uniform distribution, corresponding to pure exploration. As $\tau \to 0^+$, the distribution concentrates on the highest-scoring researchers, recovering a purely exploitative policy. Intermediate values yield a mixed strategy that optimally balances predictive performance and uncertainty under the stated objective.

Importantly, this lottery is not an *ad hoc* mechanism: it is the unique optimizer of a regularized expected-utility problem. Closely related choice rules appear in discrete choice theory, auction design, portfolio selection, and reinforcement learning.

**Cardinality constraint and operational setting.** In practical funding programs, a finite number of projects is supported in each call. To reflect this operational constraint, we impose a cardinality limit by selecting exactly $K$ researchers per funding round. This constraint ensures feasibility and aligns the allocation mechanism with the discrete nature of real-world funding decisions.

The stochastic policy derived above defines selection probabilities $\{p_i\}$ for each researcher. In practice, these probabilities are implemented by sampling $K$ researchers *without replacement* according to $\{p_i\}$. Over repeated funding rounds, the expected frequency with which each researcher is selected converges to $p_i$, thereby preserving the theoretical properties of the optimal stochastic policy while respecting the finite-capacity constraint.

**Two-round allocation and seed grants.** Once the $K$ funded researchers are selected, budget allocation proceeds in two stages. In the first stage, each selected researcher receives a fixed seed grant of size $g$, which represents a minimum level of support. The seed grant ensures that the exploration component of the stochastic policy translates into substantive funding rather than merely symbolic selection and that a diverse set of projects can be initiated. This design decouples selection from subsequent resource allocation, allowing exploration to be explicitly implemented at project initiation. The constraint $Kg \leq B$ guarantees that the total amount devoted to seed funding does not exceed the available budget.

**Exploitation among selected researchers.** After the allocation of seed grants, the remaining budget

$$B_{\text{rem}} = B - Kg$$

is distributed among the $K$ selected researchers in a second round. Let $\mathcal{S}$ denote the set of selected researchers. Each researcher $i \in \mathcal{S}$ receives an additional allocation as in Eq. 1. The parameter $\gamma$ explicitly controls how the remaining budget is split among the selected researchers. When $\gamma = 1$, the allocation is proportional to predicted performance, yielding a linear exploitation rule. For $\gamma > 1$, the allocation becomes increasingly concentrated on the highest-scoring researchers within the selected set, amplifying differences in predicted

9

performance. Conversely, for $0 < \gamma < 1$, the allocation is more egalitarian, compressing score differences and distributing resources more evenly among selected researchers. In the limit $\gamma \to \infty$, the entire remaining budget is allocated to the highest-scoring researcher among those selected.

This second-stage allocation exploits predictive information conditional on selection, while the stochastic selection stage preserves exposure to low-probability, high-impact outcomes.

# 2 Results

## 2.1 Data collection and cohort description

The empirical analysis is based on a cohort of researchers affiliated with a set of large, scientifically homogeneous research institutions in Spain. Starting from the complete list of Spanish institutions indexed in OpenAlex, we selected the 20 largest institutions satisfying the homogeneity criterion described in the Methods section. These institutions are thematically focused research centers rather than comprehensive universities, and collectively span a wide range of scientific domains, including biomedical research, oncology, astrophysics, geology, philosophy, materials science, environmental sciences, medicine, and public health. Although heterogeneous at the national level, each institution is internally coherent in terms of research themes, publication practices, and citation dynamics, making them suitable units for within-group normalization and comparative analysis.

For each institution, we identified all researchers whose current or most recent affiliation in OpenAlex corresponded to that institution and whose affiliation history indicated active association during the reference period. For each researcher, bibliometric indicators were computed over two consecutive five-year windows: the reference period $[Y-5, Y-1]$ and the future period $[Y, Y+4]$, as described in the Methods section.

After applying the inclusion criterion of having more than one publication during the reference period, the final dataset comprises approximately 4 400 researchers across the 20 institutions. This filtering step removes cases dominated by sparse or incidental publication activity and ensures that all included researchers have a minimally observable research trajectory in the reference period.

The resulting cohort exhibits substantial diversity in productivity and citation impact, both within and across institutions, reflecting the known heterogeneity of scientific performance even within narrowly defined research domains. At the same time, the use of institution-level normalization ensures that comparisons are based on relative standing within homogeneous scientific contexts, rather than on absolute bibliometric counts that are strongly field-dependent.

This dataset serves as the empirical basis for subsequent analyses of predictability, including exploratory visualization, cross-period correlation analysis, dimensionality reduction, and rank-based regression modeling.

## 2.2 Exploratory data analysis

We begin the empirical analysis with an exploratory characterization of the relationships between bibliometric indicators measured in the reference period and their counterparts in the future period. Given the highly skewed nature of scientific output and impact, this step is essential for identifying dominant regimes, non-linear patterns, and structural heterogeneity before formal modeling.

**Joint distributions across periods.** Figure 1 shows smoothed two-dimensional histograms of normalized indicators in the future period as a function of their values in the reference period. For all three measures—productivity, average citation impact, and maximum citation impact—the joint distributions exhibit a clear monotonic trend, indicating that relative standing in the reference period is informative of future performance. However, the relationship is far from homogeneous. In all cases, the density concentrates along a diagonal ridge at high values, while a broad and diffuse region is observed at lower values.

This structure reveals at least two distinct regimes. One regime corresponds to researchers with sustained high productivity or impact, for whom future performance scales approximately monotonically with past performance. The second regime comprises researchers with low or moderate activity, for whom future outcomes display substantially higher variability. This separation is particularly pronounced for productivity and maximum citation impact and is consistent with the intuition that stable research pipelines coexist with more sporadic scientific output.

**Heavy-tailed behavior and regime separation.** The heavy-tailed nature of the bibliometric indicators is further illustrated by the complementary cumulative distribution functions (CCDFs) shown in Figure 2. When plotted on logarithmic scales, the distributions of productivity, average citation impact, and maximum citation impact in the future period display long tails spanning several orders of magnitude. These tails indicate that a small fraction of researchers accounts for a disproportionate share of output and impact.

Notably, the CCDFs reveal a marked curvature separating low-activity and high-activity regimes, particularly for productivity and maximum citation impact. This observation is consistent with the joint distributions and suggests that a single characteristic scale cannot adequately summarize scientific performance. Instead, the data support a heterogeneous picture in which distinct performance regimes coexist within the same scientific context.

**Conditional distributions for high-impact researchers.** To further examine the characteristics of researchers associated with high-impact outcomes, we focus on the subset of researchers whose maximum future citation impact lies in the upper half of the empirical distribution. For this subset, we compute the empirical cumulative distribution functions of productivity and average citation impact in the future period (Figure 3).

11

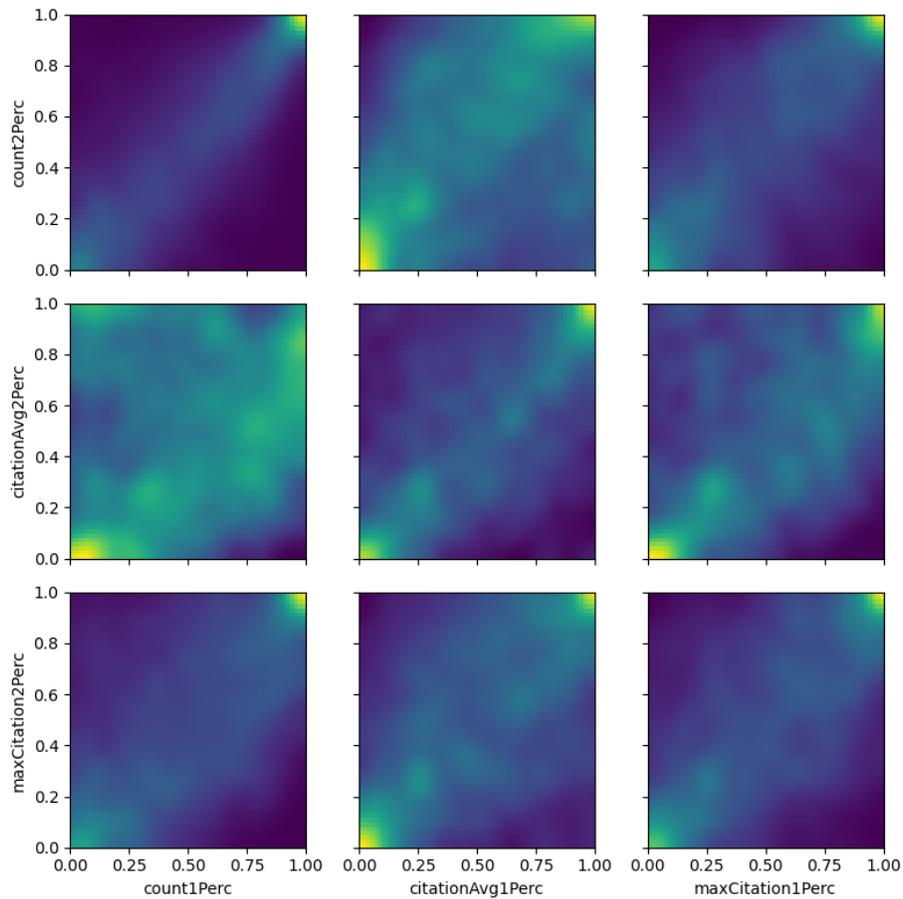

Figure 1: Smoothed two-dimensional histograms showing the relationship between percentile-normalized bibliometric indicators measured in the reference period (horizontal axes) and the future period (vertical axes). Rows correspond to productivity, average citation impact, and maximum citation impact in the future period, while columns correspond to the same indicators in the reference period. The density structure reveals distinct performance regimes and nonlinear persistence patterns across periods.

12

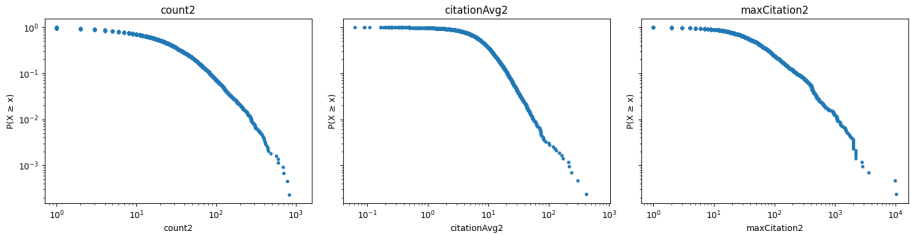

Figure 2: Complementary cumulative distribution functions (CCDFs) of future-period productivity, average citation impact, and maximum citation impact plotted on logarithmic scales. All three distributions exhibit heavy-tailed behavior spanning multiple orders of magnitude, with particularly pronounced tails for maximum citation impact. These patterns indicate that a small fraction of researchers accounts for a disproportionate share of scientific output and influence.

Despite conditioning on high maximum citation impact, substantial heterogeneity in scientific activity remains. Focusing on researchers whose maximum citation impact in the future period lies above the median, the median number of publications over the five-year window is 38, corresponding to approximately eight publications per year. By construction, this implies that half of the researchers in this high-impact subset publish more than 38 papers during the period, while the other half produce fewer. At the same time, the median average number of citations per publication for this group is 11.25. These values indicate that highly influential contributions are not confined to extreme levels of productivity or average citation impact. In particular, exceptional papers may arise from both researchers with sustained high publication volumes and those with more moderate output, reinforcing the notion that breakthrough contributions are only partially predictable from aggregate performance measures.

**Correlation structure across periods.** We examine the correlation structure between percentile-normalized indicators measured in the reference and future periods to quantify temporal persistence in scientific performance. Table 1 reports consistently positive cross-period correlations for all indicators, confirming that past performance contains predictive information about future outcomes. However, the strength of this predictability varies markedly across measures. Productivity exhibits the strongest temporal persistence, with rank correlations exceeding 0.7 between consecutive five-year windows, indicating that relative publication output is highly stable over time. In contrast, citation-based indicators show substantially weaker correlations, reflecting greater variability and sensitivity to stochastic effects. Correlations involving maximum citation impact are intermediate in magnitude, suggesting that while researchers who have produced highly influential contributions in the past are more likely to do so again, exceptional impact remains only partially predictable. Composite

13

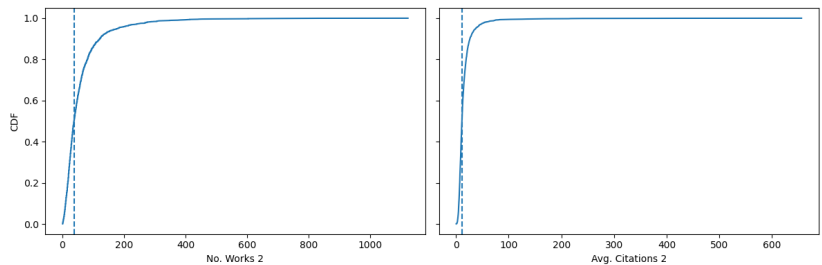

Figure 3: Empirical cumulative distribution functions (CDFs) of productivity (*left*) and average citation impact (*right*) in the future period, conditional on researchers belonging to the upper half of the distribution of maximum citation impact. Vertical dashed lines indicate the median values for each distribution. The results highlight substantial heterogeneity in productivity and average impact even among researchers associated with highly influential contributions.

Table 1: Cross-period rank correlations between reference-period indicators (rows) and future-period indicators (columns). All values correspond to Pearson correlations computed on percentile-normalized ranks within homogeneous scientific domains.

|                     | count2Perc | citationAvg2Perc | maxCitation2Perc | avgPerc2 |
|---------------------|------------|------------------|------------------|----------|
| **count1Perc**      | 0.747      | 0.117            | 0.497            | 0.681    |
| **citationAvg1Perc**| 0.300      | 0.290            | 0.362            | 0.439    |
| **maxCitation1Perc**| 0.539      | 0.250            | 0.482            | 0.601    |
| **avgPerc1**        | 0.605      | 0.251            | 0.511            | 0.656    |

indicators, such as the average percentile rank across measures, display relatively strong correlations with future performance, indicating that aggregating multiple dimensions of past performance captures more stable information than any single metric alone. Taken together, these results reinforce three key empirical features identified in the exploratory analysis: the coexistence of distinct performance regimes, the heavy-tailed nature of scientific output and impact, and the presence of partial but imperfect predictability across periods. These observations motivate the rank-based and probabilistic modeling approaches developed in the following sections.

### 2.3 Dimensionality reduction and collinearity analysis

To assess redundancy among reference-period bibliometric indicators and identify a parsimonious representation of past performance, we conducted a Principal Component Analysis (PCA) and a collinearity assessment using the Variance Inflation Factor (VIF).

PCA was applied to the set of percentile-normalized reference-period indicators, namely productivity, average citation impact, and maximum citation

14

impact. The results show an intense concentration of variance in the first principal component. Specifically, the first principal component (PC1) accounts for approximately 77% of the total variance, while the second principal component (PC2) accounts for an additional 21%. Together, the first two components account for more than 97% of the total variance, with the third component contributing only marginally (approximately 2%). This rapid decay in explained variance indicates that a low-dimensional representation can capture most of the information contained in the three original indicators.

The factor loadings provide further insight into the structure of this reduced space. All three indicators load positively and substantially on PC1, with substantial contributions from maximum and average citation impact, indicating that this component captures a general dimension of overall past performance. The second component contrasts productivity against citation-based measures, reflecting a trade-off between publication volume and per-paper impact. The third component, which explains only a negligible fraction of the variance, is dominated by opposing contributions of maximum citation impact and the remaining indicators, and does not appear to carry substantial independent information.

Complementary to the PCA, we assessed multicollinearity using the Variance Inflation Factor. When all three reference-period indicators are included simultaneously, VIF values exceed the commonly used threshold of 5, indicating substantial collinearity and limited marginal explanatory power for at least one of the variables. In contrast, when restricting the model to productivity and maximum citation impact, both variables exhibit VIF values below 2, indicating low collinearity and stable estimation properties.

Taken together, the PCA and VIF analyses lead to a consistent conclusion: while productivity, average citation impact, and maximum citation impact all capture meaningful aspects of past performance, a two-variable representation based on productivity and maximum citation impact retains most of the informative content while avoiding excessive collinearity. This parsimonious representation is therefore adopted in the subsequent predictive and allocation models.

## 2.4 Rank-based regression analysis

To quantify the predictive relationships between reference-period performance and future outcomes, we fitted a series of rank-based regression models in which future-period percentile-normalized indicators were regressed on reference-period percentile-normalized indicators. All models were estimated using rank-transformed variables, thereby focusing on relative standing rather than absolute magnitudes and ensuring robustness to heavy-tailed distributions and extreme observations.

We first considered models including two reference-period predictors: productivity and maximum citation impact. Across all targets, these models explain a non-negligible fraction of the variance in future performance, with explanatory power varying by indicator. Future productivity is the most predictable outcome, with a coefficient of determination exceeding 0.55. Future

15

maximum citation impact and the composite performance indicator exhibit intermediate predictability, while future average citation impact shows substantially lower predictability, consistent with the exploratory analysis.

An examination of the regression coefficients and partial $R^2$ values reveals a clear hierarchy among predictors. Productivity in the reference period consistently makes the most significant contribution to the explained variance. For future productivity, reference-period productivity alone accounts for more than 26% of the variance explained by the complete two-variable model, whereas maximum citation impact contributes negligibly. Similarly, for the composite performance indicator, the contribution of productivity is substantially larger than that of maximum citation impact. For future maximum citation impact, both predictors contribute meaningfully, but productivity again explains a slightly larger share of the variance.

In contrast, future average citation impact shows little overall predictability. Although maximum citation impact in the reference period contributes more than productivity in relative terms, the total variance explained remains small, underscoring the limited ability of past indicators to forecast per-paper citation intensity.

We then considered single-predictor models to assess which reference-period indicator performs best in isolation. In all cases, productivity in the reference period emerges as the strongest single predictor. For future productivity, the rank regression on reference-period productivity alone explains approximately 56% of the variance, nearly matching the explanatory power of the two-variable model. For future maximum citation impact and the composite indicator, productivity alone outperforms maximum citation impact alone in terms of explained variance. Even for future average citation impact, where overall predictability is low, productivity provides a non-negligible signal.

Taken together, these results provide empirical support for a core assumption underlying the often-cited "publish or perish" principle. Beyond the well-known social and institutional pressure to publish, the publication rate in the reference period emerges as a robust predictor of future scientific performance. In particular, reference-period productivity is highly predictive not only of future productivity but also of future citation-based impact and exceptional contributions, outperforming citation-based indicators when considered in isolation. This finding suggests that sustained publication activity captures a combination of factors—including research continuity, integration in active scientific networks, and the maintenance of productive research pipelines—that are directly relevant for future output and influence. Importantly, this result does not imply that publication quantity should be maximized without regard to quality, but rather that productivity constitutes a reliable summary signal of future scientific activity under conditions of heavy-tailed uncertainty. At the same time, the incomplete predictability of citation-based outcomes highlights the limits of purely deterministic funding strategies. It motivates the inclusion of uncertainty-aware mechanisms in research funding allocation.

## 2.5 Aggregated past performance as a proxy signal.

While individual indicators capture complementary aspects of scientific performance, funding decisions typically rely on aggregate assessments rather than on single metrics. To this end, we define a composite proxy signal, `avgPerc1`, obtained by averaging the three percentile-normalized indicators measured in the reference period: productivity, average citation impact, and maximum citation impact. By construction, this aggregated signal integrates sustained output, typical influence, and exceptional contributions into a single, field-normalized measure.

Figure 4 illustrates the relationship between this proxy signal and future performance. Each panel shows boxplots of a percentile-normalized outcome in the future period—productivity, average citation impact, and maximum citation impact—conditioned on deciles of `avgPerc1`. In all three cases, a clear monotonic trend is observed: higher values of the aggregated past-performance signal are associated with systematically higher medians and upper quantiles of future outcomes. This pattern provides a robust, nonparametric visualization of predictability that is consistent across all performance dimensions.

At the same time, substantial dispersion is present within each decile, particularly for citation-based measures. Even among researchers with low aggregated past performance, the distributions exhibit long upper tails, indicating that highly influential or productive outcomes can still occur. Conversely, researchers in the highest deciles of `avgPerc1` display markedly higher probabilities of strong future performance, but with considerable variability that cannot be eliminated by aggregation. These results underscore a central empirical finding of this study: combining multiple past-performance indicators improves predictive power at the population level, yet irreducible uncertainty persists at the individual level, especially for exceptional outcomes.

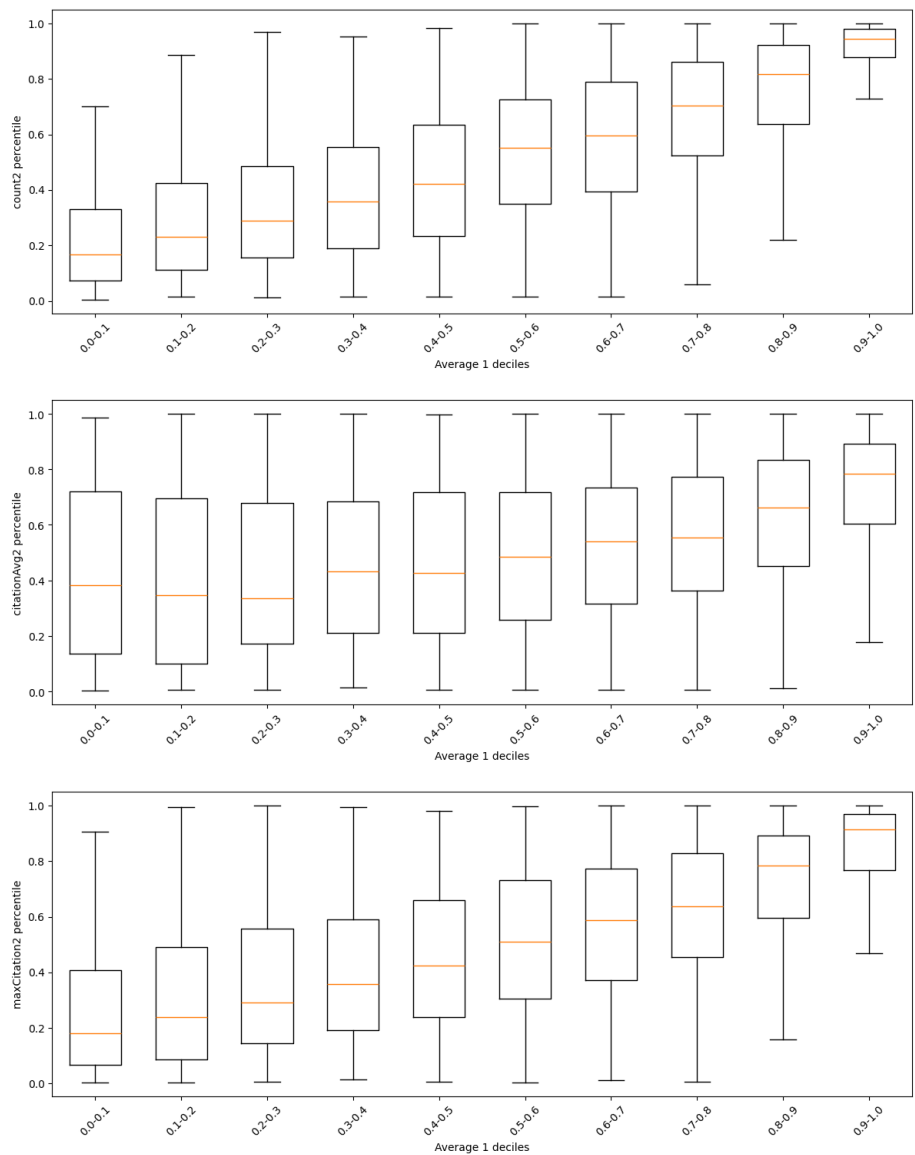

Figure 4: Future performance as a function of aggregated past performance. Boxplots show percentile-normalized outcomes in the future period—productivity (top), average citation impact (middle), and maximum citation impact (bottom)—grouped by deciles of the aggregated past-performance signal `avgPerc1`. Across all outcomes, higher values of `avgPerc1` are associated with higher median and upper-quantile performance, while substantial dispersion within each decile highlights the persistence of uncertainty and heavy-tailed effects.

18



\end{document}